\documentclass[prb,showpacs]{revtex4} % <== Physical Review B
\usepackage{graphicx}                 % <== Include figure files
\begin{document}

\title{Electronic band structure and chemical bonding\\
in the novel antiperovskite ZnCNi$_{3}$ as compared\\
with 8-K superconductor MgCNi$_{3}$.}
\author{I.R. Shein$^*$, K.I.Shein, N.I.Medvedeva and A.L. Ivanovskii}

\affiliation {Institute of Solid State Chemistry, Ural Branch of
the Russian Academy of Sciences, 620219, Ekaterinburg, Russia}

\begin{abstract}
Energy band structure of the discovered ternary perovskite-like
compound ZnCNi$_{3}$ reported by Park et al (2004) as a
non-superconducting paramagnetic metal was investigated using the
FLMTO-GGA method. The electronic bands, density of states, Fermi
surface, charge density and electron localization function
distributions for ZnCNi$_{3}$ are obtained and analyzed in
comparison with the isoelectronic and isostructural 8K
superconductor MgCNi$_{3}$. The effect of external pressure on
the electronic states of ZnCNi$_{3}$ and MgCNi$_{3}$ is studied.\\

$^*$ E-mail: shein@ihim.uran.ru
\end{abstract}


% 74.70.-b Superconducting materials (excluding high-Tc compounds)
% 71.20.Dg Alkali and alkaline earth metals
% 71.20.-b Electron density of states and band structure of
%          crystalline solids

\maketitle The interest in the properties of the perovskite - like
intermetallic systems has increased in the past years due to the
recent discovery of superconductivity (SC) with transition
temperature up to 8 K in the perovskite - like MgCNi$_{3}$
by\cite{He}. Several factors attach special significance to this
discovery. Among perovskite - like phases, the MgCNi$_{3}$ is the
first oxygen-free superconductor. Among low-temperature
superconductors, the MgCNi$_{3}$ also occupies a special place.
Because large content of Ni in MgCNi$_{3}$, it is expected that
this material is in close proximity to ferromagnetism. The
occurrence of superconductivity in this system is very unusual;
this makes it similar to the discovered recently ferromagnetic
superconductors, and MgCNi$_{3}$ can be treated as an
"intermediate phase" between the groups of conventional
nonmagnetic and ferromagnetic superconductors. Since 2001, the
properties of the MgCNi$_{3}$ have been investigated in detail,
see review\cite{Ivanovskii}. However till now the nature of the
pairing mechanism in this SC material is still controversial. On
the one hand, the temperature dependence of specific heat, the NMR
data and some theoretical estimations allow to suggest that
MgCNi${_3}$ is a classical superconductor with isotropic s-wave
BCS pairing. On the other hand, a microwave impedance and
tunneling data can be considered as arguments of non-s-wave behavior.\\
The great attention has been paid to influence of the doping
effects on MgCNi$_{3}$ properties, and various solid solutions
such as MgCNi$_{3-x}$M$_{z}$ (M = d-metals) are synthesized.\\
Quite recently, Park et al\cite{Park} have carried out a synthesis
of novel ZnCNi$_{3}$ phase by solid-state route and characterized
by X-ray powder diffraction (XRD), dc-resistivity, magnetization
and specific heat measurements. It was found that ZnCNi$_{3}$ is
isostructural with 8Ê SC MgCNi$_{3}$ however remains a
paramagnetic metal up to T $<$ 2 K. The absence of
superconductivity in ZnCNi$_{3}$ is interpreted \cite{Park} within
the BCS framework in terms of the reduction in lattice constant a
(about 3.6 \%) will lead to sharply decrease in the total density
of states at the Fermi level N(E${_F}$).\\
In this report, we are focusing on the electronic band structure
and chemical bonding in ZnCNi$_{3}$ in comparison with MgCNi$_{3}$
by means of a scalar relativistic full-potential linear muffin-tin
method (FLMTO) with the generalized gradient approximation (GGA)
for correlation and exchange effects \cite{Savrasov,Perdew}. In
order to analyze the bonding effects, we have calculated the
electronic localization function (ELF \cite{Becke,Savin}) and
charge density ($\rho$) distributions. The interatomic bond
indices (crystal orbital overlap populations - COOPs) were also
estimated using simple band structure thigh-binding EHT
approach\cite{Hoffmann}.\\
ZnCNi$_{3}$ has the cubic perovskite-like structure (space group
Pm3m) consisting of Zn at the corners, C at the body center, and
Ni at the face centers of the cube. The atomic positions are Ni:
3ñ (0.5,0.5,0); Zn: 1a (0,0,0); C: 1b (0.5,0.5,0.5). At the first
stage the lattice parameter of ZnCNi$_{3}$ has been optimized
(a=0.3747 nm). We find that this value is about 2.3\%
overestimated as compared with experiment (0.366 nm \cite{Park})
as is typical of generalized gradient calculations within the LDA.\\

\begin{figure*}[!htb]
\vskip  0cm
\begin{tabular}{c}

\includegraphics[width=16.0 cm,clip]{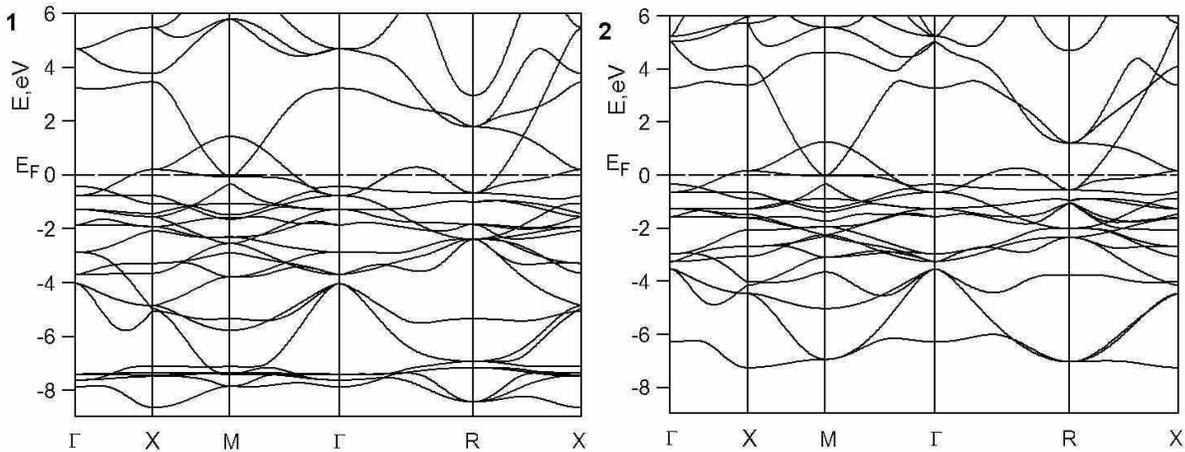}\\
\end{tabular}
\vspace{-0.02cm} \caption[a] {\small. Energy bands for ZnCNi$_{3}$
(1) and MgCNi$_{3}$ (2).}
%\end{floatingfigure}
\end{figure*}

\begin{figure*}[!htb]
\vskip  0cm
\begin{tabular}{c}

\includegraphics[width=16.0 cm,clip]{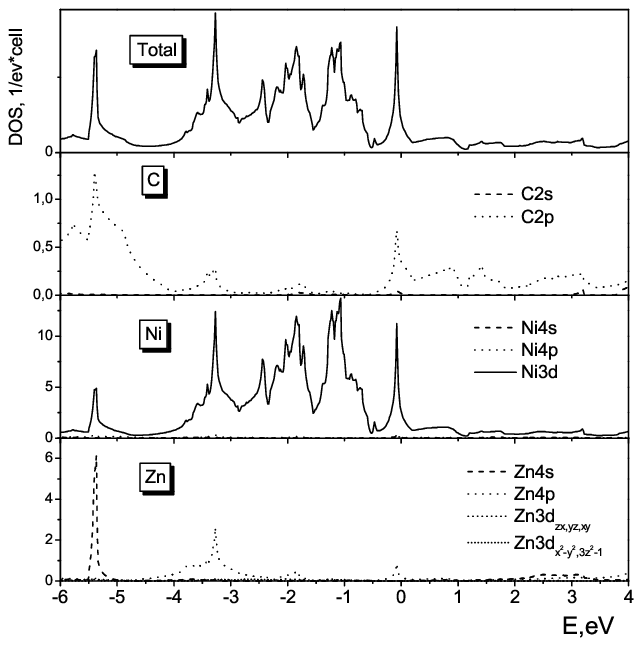}\\

\end{tabular}

\vspace{-0.02cm} \caption[a] { \small . Total and site- projected
$\emph{l}$ - decomposed densities of states for ZnCNi$_{3}$.}
%\end{floatingfigure}
\end{figure*}

The energy bands, total and site-projected \emph{l}-decomposed
densities of states (DOS) for ZnCNi${_3}$ and MgCNi${_3}$ are
shown in Figs. 1 and 2, respectively. The valence band in
ZnCNi${_3}$ is derived basically from the 15 Ni3d and 3 C2p states
filled by 34 electrons (Fig. 1). C2s and Zn states play a minor
role in this area. The C 2p states are hybridized with Ni3d, and
are located below nine near-Fermi bands which are formed mainly
from Ni3d states. From these bands three are bonding, three
nonbonding, and three antibonding bands. Two of antibonding
$\pi$-bands cross E${_F}$. The upper band produces hole-like two -
sheet Fermi surface centered at the X point and ellipse-like along
the $\Gamma$-R line, Fig 3. The Fermi surface involves also
electron-like sheets around $\Gamma$ point and along the R-M line
generated by the lower band. As is seen from Fig. 1, energy bands
of isostructural and isoelectronic phase MgCNi$_{3}$ are quite
similar.\\

\begin{figure*}[!htb]
\vskip  0cm
\begin{tabular}{c}

\includegraphics[width=10.0 cm,clip]{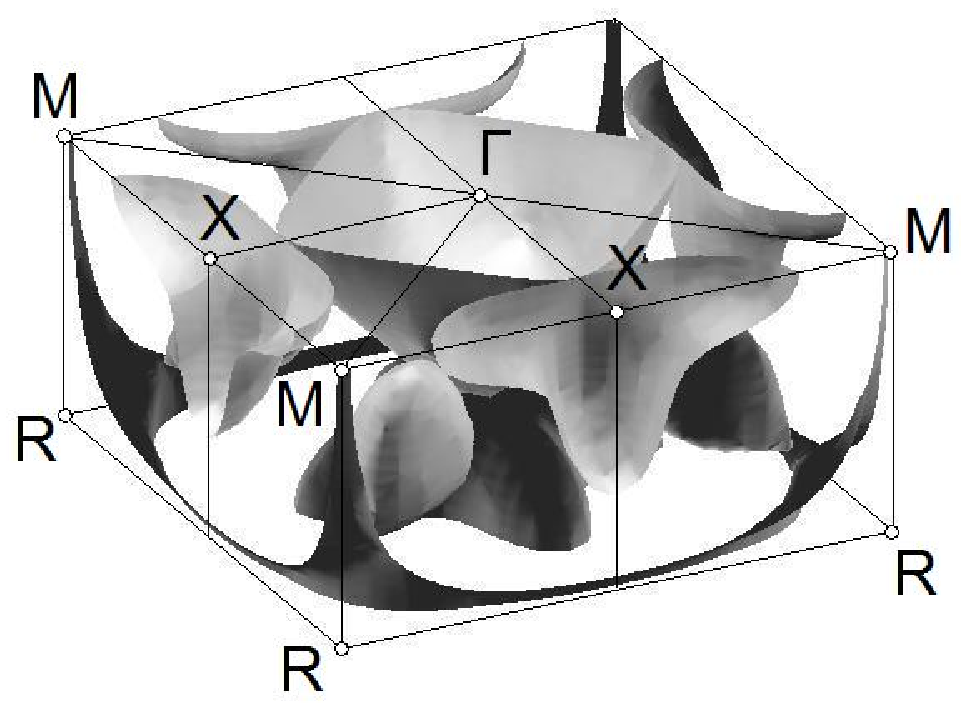}\\

\end{tabular}

\vspace{-0.02cm} \caption[a] { \small Fermi surface for
ZnCNi$_{3.}$}
%\end{floatingfigure}
\end{figure*}

The most remarkable feature of the DOS is a narrow intense peak in
the vicinity of the Fermi level (Fig. 2). This peak is associated
with the quasi-flat Ni 3d band aligned along the X-M and
M-$\Gamma$ directions in the Brillouin zone. The Fermi level is
located at the high-energy slope of the above peak. The maximum
contribution (~ 84.4 \%, Table 1) to the density of states at the
Fermi level N(E$_{F}$ is made by the Ni 3d states, and these
states are responsible for  metallic properties of the material.\\
Let's consider the chemical bonding in ZnCNi$_{3}$. The ELF
distribution for the section through Zn, C and Ni atoms in
ZnCNi$_{3}$ is presented in Fig. 4. According to spatial
organization of this scalar function $\Lambda$ \cite{Becke,Savin}
which describes the local kinetic energy, the ELF can take values
in the interval 0 $<$ 1. Here the value $\Lambda$ = 1 corresponds
to perfect localization and $\Lambda$ = 0.5 to the case of a free
electron gas. For ZnCNi$_{3}$ a maximum value of the $\Lambda$ in
the range about 0.8 around the carbons indicate an anionic state
of these atoms in the crystal. The presence of a local minimum of
the $\Lambda$ on the line connecting Ni and C atoms may be
interpreted as covalent bonding. Metallic bonding, which is an
intermediate case between the covalent and the ionic bonding, is
described by a rather uniform $\Lambda$ distribution in region
between Ni and Zn atoms. Corresponding electron density
distributions $\rho$ can be well distinguished on the charge map, Fig. 4.\\

\begin{figure*}[!htb]
\vskip  0cm
\begin{tabular}{c}

\includegraphics[width=16.0 cm,clip]{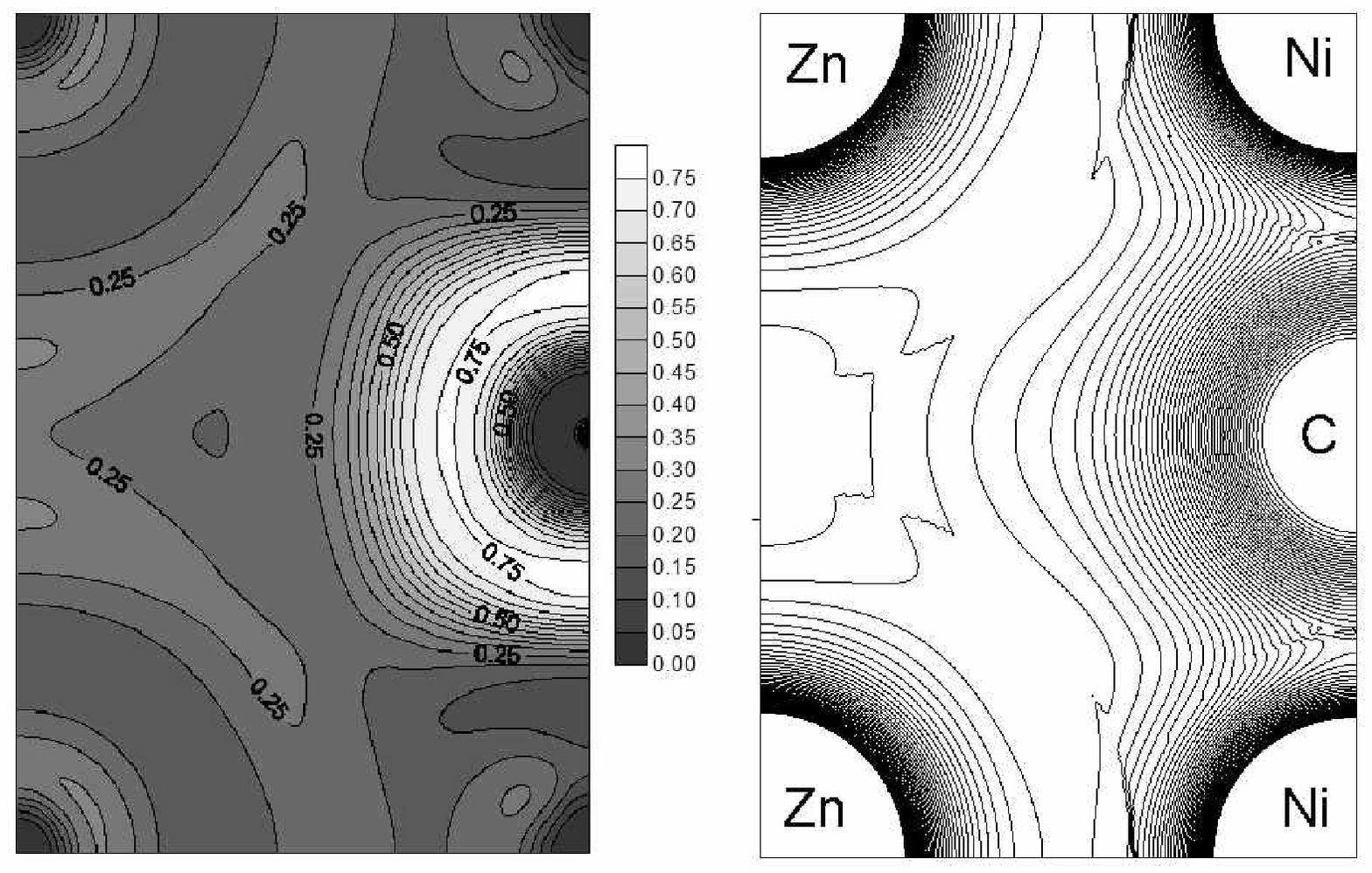}\\

\end{tabular}

\vspace{-0.02cm} \caption[a] { \small ELF (left) and charge
density contours (right) for the (110) section of ZnCNi$_{3}$.}
%\end{floatingfigure}
\end{figure*}

In order to analyze in detail the bonding picture in ZnCNi$_{3}$
we have calculated the crystal orbital overlap populations (COOPs)
values by means of simple tight binding band structure method
within the extended Huckel theory (EHT)
approximation\cite{Hoffmann}. The results obtained indicate (Table
2) that the Ni-C covalent interactions are  the strongest bonds in
ZnCNi$_{3}$, whereas other bond types (Ni-Zn and Zn-C) are
appeared on the order less. Interestingly, that the Ni-C COOP
values in ZnCNi$_{3}$ are about 3.4\% weaker as compared with
MgCNi$_{3}$, in spite of the fact that Ni-C distances are shorter.
On the contrary, the degree of covalency of Ni-Zn and Zn-C bonds
in ZnCNi$_{3}$ as compared with Ni-Mg and Mg-C bonds in MgCNi$_{3}$ grows.\\
Let's address to relation between electronic structure and
superconductivity. As is known, in the BCS strong coupling limit,
T$_{c}$  is expressed by the McMillan formula\cite{McMillan}:
T$_{c}$ $\sim$ $<\omega>$ exp {f($\lambda$)}, where  $<\omega>$
represents an averaged phonon frequency, the coupling constant
$\lambda$ = N(E$_{F}$)$<I^{2}>$/M$<\omega^{2}>$, $<I^{2}>$ -
frequency is an averaged electron-ion matrix element squared, M is
an atomic mass, and N(E$_F$) is the density of states at the Fermi level.\\
Our FLMTO-GGA calculations showned, that the N(EF) values for
MgCNi$_{3}$ (5.399) and  ZnCNi$_{3}$ (4.401 states/eV) differ
less, than on ~ 18.5 \%. Thus, our data sharply differ from crude
estimations of authors \cite{Park}, according to which the absence
of superconductivity in ZrCNi$_{3}$ can be achieved if the ratio
N(E$_{F}$)[ZnCNi$_{3}$]/N(E$_{F}$)[MgCNi$_{3}$]$\sim$ 0.41. As one
of the possible reasons of rapid reduction of N(E$_{F}$) the
proximity of the Fermi energy to a large near-Fermi peak in the
DOS (van Hove singularity) is assumed \cite{Park}. It was
supposed, as a decreases, the broadening of van Hove
singularity peak is accompanied by the fast decreasing in N(E$_{F}$).\\
To check this assumption we have analyzed the pressure dependence
of the near-Fermi DOSs for MgCNi$_{3}$ and ZnCNi$_{3}$. The
results obtained indicate (Table 1) that with growth of external
pressure (reduction in lattice parameter à) the N(E$_F$) both for
MgCNi$_{3}$ and ZnCNi$_{3}$ decreases monotonically, however the
ratio N(E$_{F}$)[ZnCNi$_{3}$]/N(E$_{F}$)[MgCNi$_{3}$] varies a
little. For MgCNi$_{3}$ these results are in agreement with
earlier theoretical and experimental findings \cite{Kumary},
according to which external pressure leads to T$_{c}$ reduction
as result of N(E$_{F}$) decreasing.\\
In conclusion, we have presented the first principle band
structure calculations for the newly discovered ternary
intermetallic perovskite-like ZnCNi$_{3}$ in comparison with
isoelectronic and isostructural 8Ê superconductor MgCNi$_{3}$
performed by the FLMTO-GGA method. It was established that the
general features of ZnCNi$_{3}$ and MgCNi$_{3}$ band structure and
interatomic bonding are similar and can be briefly summarized as
follows: there is a strong hybridization between the Ni-3d
electrons and the C-2p states and thus carbon plays the crucial
role of the mediator of electron hopping. Two of the (Ni3d -
C2p)-like bands cross the Fermi level and contribute to
conductivity. Remarkable feature of the DOS of these materials is
a sharp peak below E$_{F}$ arising from the $\pi$ - antibonding
Ni3d states. The N(E$_F$) is formed predominantly due to Ni3d
states. According to ELF and charge density analysis, the bonding
in ZnCNi$_{3}$ is of "mixed" character: the ionic contribution
(due to additional charge localization on C atoms) in this phase
coexists with Ni-C covalent bonding  and a metallicity provided by
the delocalized Ni states. The simple TB estimations showed also
that the replacement Mg $\rightarrow>$ Zn in perovskite-like phase
results in the population redistribution between separate bond
types thus the strongest Ni-C bonds in ZnCNi$_{3}$ as compared
with MgCNi$_{3}$ are weaker in spite of the fact that Ni-C
distances are shorter.\\
Our LMTO-GGA band structure calculations show, that to explain the
absence of superconductivity in ZnCNi3 based only on the
electronic factor (reduction in N(E$_{F}$) as compared with
MgCNi$_{3}$ \cite{Park} is impossible. Obviously, the important
role in the observed non-superconducting state in ZnCNi$_{3}$ will
belong to the changes in phonon frequencies as well as to possible
structural and chemical defects. For example, the effect of Ñ
vacancies on the value of T$_{c}$ is noticeable for MgCNi$_{3}$.\\

Acknowledgement.\\
This work was supported by the RFBR, grants 02-03-32971 and
04-03-32082.

\begin{center}
\begin{table}
\caption{Total N(E$_{F}$) and site- projected l- decomposed
densities of states at the Fermi level  (states/eV cell) as
function of external pressure (P, GPa) for antiperovskites
ZnCNi$_{3}$ and MgCNi$_{3}$.}
\begin{tabular}{|c|c|c|c|c|c|c|c|c|c|c|c|}
\hline
P&293.7&232.6&182.0&140.2&105.7&77.5&54.4&35.6&20.3&8.1&0\\
\hline
C2s&0.019&0.019&0.019&0.019&0.019&0.019&0.020&0.020&0.020&0.020&0.021\\
\hline
C2p&0.306&0.309&0.312&0.314&0.318&0.323&0.328&0.333&0.337&0.342&0.345\\
\hline
Ni4s&0.055&0.056&0.056&0.056&0.056&0.056&0.057&0.058&0.059&0.060&0.060\\
\hline
Ni4p&0.097&0.095&0.094&0.092&0.091&0.089&0.089&0.087&0.086&0.085&0.085\\
\hline
Ni4d&2.907&2.985&3.064&3.095&3.177&3.255&3.355&3.456&3.539&3.644&3.716\\
\hline
Zn4s&0.003&0.003&0.003&0.003&0.003&0.003&0.003&0.003&0.003&0.003&0.003\\
\hline
Zn4p&0.163&0.162&0.162&0.158&0.157&0.155&0.145&0.151&0.153&0.153&0.153\\
\hline
Zn3d&0.025&0.024&0.024&0.023&0.022&0.022&0.021&0.020&0.020&0.019&0.019\\
\hline
{\bf N(E$_{F}$)}&{\bf 3.575}&{\bf 3.654}&{\bf 3.734}&{\bf
3.758}&{\bf 3.843}&{\bf 3.923}&{\bf 4.027}&{\bf 4.128}&{\bf
4.217}&{\bf 4.325}&{\bf 4.401}\\
\hline \hline
P&279.4&223.4&176.8&138.1&106.1&79.54&57.7&39.74&25.1&13.2&0\\
\hline
C2s&0.019&0.019&0.019&0.019&0.019&0.019&0.019&0.019&0.019&0.019&0.020\\
\hline
C2p&0.377&0.379&0.379&0.382&0.384&0.387&0.390&0.392&0.394&0.395&0.403\\
\hline
Ni4s&0.084&0.083&0.080&0.080&0.079&0.080&0.080&0.081&0.081&0.080&0.083\\
\hline
Ni4p&0.120&0.118&0.114&0.112&0.110&0.109&0.108&0.106&0.104&0.101&0.099\\
\hline
Ni4d&3.683&3.746&3.731&3.804&3.882&4.005&4.108&4.215&4.315&4.367&4.607\\
\hline
Mg3s&0.001&0.001&0.001&0.001&0.001&0.001&0.001&0.001&0.001&0.001&0.000\\
\hline
Mg3p&0.190&0.186&0.177&0.173&0.172&0.174&0.175&0.173&0.172&0.167&0.165\\
\hline
Mg3d&0.032&0.031&0.030&0.028&0.027&0.027&0.026&0.025&0.024&0.023&0.022\\
\hline {\bf N(E$_{F}$)}&{\bf 4.507}&{\bf 4.563}&{\bf 4.530}&{\bf
4.598}&{\bf 4.675}&{\bf 4.803}&{\bf 4.907}&{\bf 5.012}&{\bf
5.111}&{\bf 5.154}&{\bf 5.399}\\
\hline
\end{tabular}
\end{table}
\end{center}

\begin{center}
\begin{table}
\caption{The COOPs (electron/bond) of interatomic bonds in
ZnCNi$_{3}$ and MgCNi$_{3}$ according tight-binding EHT
estimations.}
\begin{tabular}{|c|c|c|c|c|c|}
\hline \multicolumn{3} {|c|} {ZnCNi$_3$}&\multicolumn{3} {|c|} {MgCNi$_3$}\\
\hline
Bonds&Distances,nm&COOP's&Bonds&Distances,nm&COOP's\\
\hline
Ni-C&0.1876&0.2898&Ni-C&0.1898&0.2999\\
\hline
Ni-Zn&0.2653&0.0744&Ni-Mg&0.2684&0.0393\\
\hline
Zn-C&0.3249&0.0056&Mg-C&0.3287&0.0017\\
\hline
\end{tabular}
\end{table}
\end{center}

%%%%%%%%%%%%%%%%%%%%%%%%%%%% Bibliography %%%%%%%%%%%%%%%%%%%%%%%%%%%

\end{document}